\documentclass[conference]{IEEEtran}
 \ifCLASSINFOpdf \else \fi

\usepackage{amssymb}
\usepackage{amsmath}
\usepackage{epsfig}

\begin{document}
\newtheorem{theorem}{Theorem}
\newtheorem{corollary}{Corollary}
\newtheorem{conjecture}{Conjecture}
\newtheorem{definition}{Definition}
\newtheorem{lemma}{Lemma}

% paper title
% can use linebreaks \\ within to get better formatting as desired
\title{Interference, Cooperation and Connectivity --  A Degrees of Freedom Perspective}

% author names and affiliations
% use a multiple column layout for up to three different
% affiliations
\author{\IEEEauthorblockN{Chenwei Wang\authorrefmark{1}, Syed A. Jafar\authorrefmark{1}, Shlomo Shamai (Shitz)\authorrefmark{2} and Michele Wigger\authorrefmark{3}}
\authorblockA{\authorrefmark{1}EECS Dept., University of California Irvine, Irvine, CA, 92697, USA}
\authorblockA{\authorrefmark{2}EE Dept., Technion-Israel Institute of Technology, Technion City, Haifa 32000, Israel}
\authorblockA{\authorrefmark{3}CE Dept., UTelecom ParisTech, Paris, 75634, France}
}
\maketitle

\begin{abstract}
%\boldmath

THIS PAPER IS ELIGIBLE FOR THE STUDENT PAPER AWARD. We explore the
interplay between interference, cooperation and connectivity in
heterogeneous wireless interference networks. Specifically, we
consider a 4-user locally-connected interference network with
pairwise clustered decoding and show that its degrees of freedom
(DoF) are bounded above by $\frac{12}{5}$. Interestingly, when
compared to the corresponding fully connected setting which is known
to have $\frac{8}{3}$ DoF, the locally connected network is only
missing interference-carrying links, but still has lower DoF, i.e.,
eliminating these interference-carrying links reduces the DoF. The
$\frac{12}{5}$ DoF outer bound is obtained through a novel approach
that translates insights from interference alignment over linear
vector spaces into corresponding sub-modularity relationships
between entropy functions.
\end{abstract}

\section{Introduction}

\allowdisplaybreaks
The broadcast nature of the wireless medium gives rise to three fundamental aspects of wireless networks.
\begin{enumerate}
\item {\it Interference -- } among concurrent transmissions. This is the greatest challenge faced by a wireless network.
\item {\it Cooperation --} among nodes, e.g., by relaying simultaneously overheard  transmissions to their desired destinations. This is the greatest opportunity present in a wireless network.
\item{\it Local Connectivity --} enforced by wireless propagation path loss, it limits the range over which signals can be heard. Therefore,  it limits both the harmful impact of interference and the benefits of cooperation.
\end{enumerate}

Understanding the complex interplay between these three factors is essential to understanding the capacity limits of wireless networks. In this work we seek to illuminate a few interesting aspects of the problem by focusing on a specific network topology that involves all three elements.

\subsection{The Problem}
\begin{figure}[!h] \vspace{-0.15in}\centering
\includegraphics[width=3.3in]{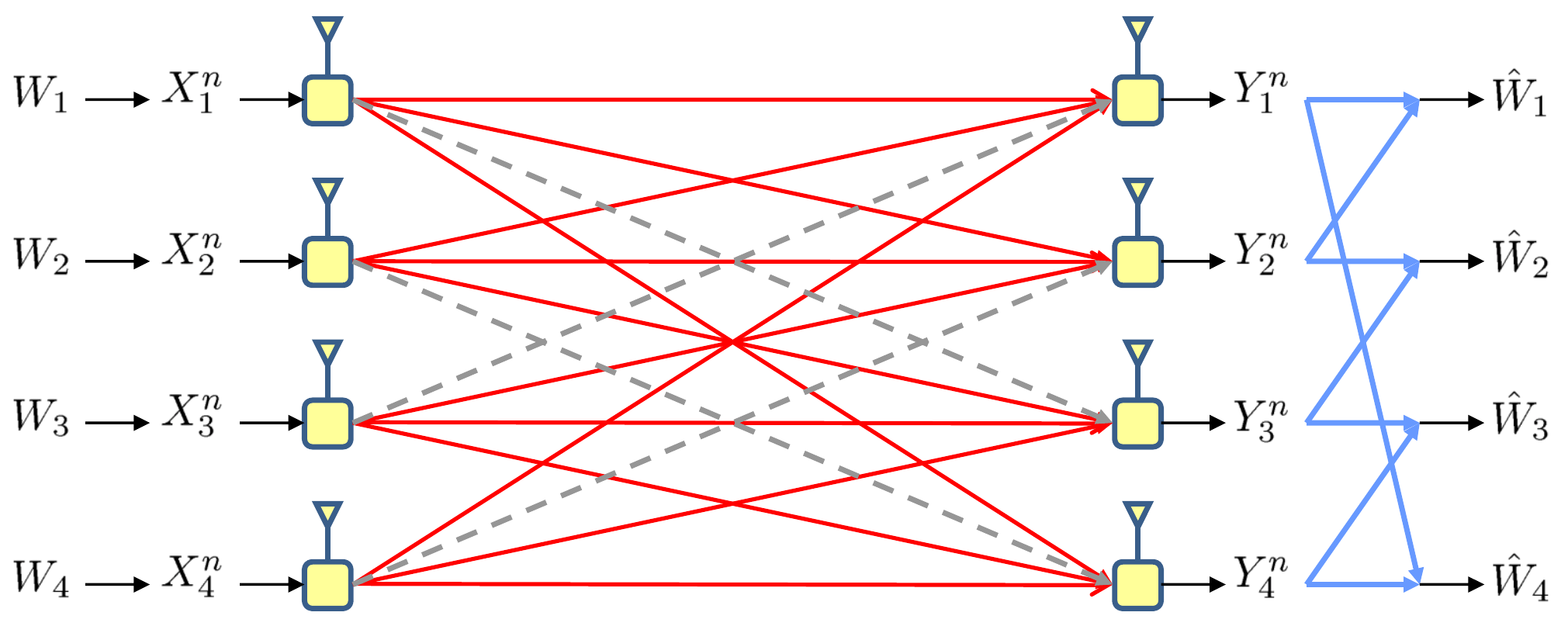}\vspace{-0.15in}
\caption{4 User Interference Channel with Pairwise Clustered
Decoding}\vspace{-0.1in} \label{fig:channelmodel}
\end{figure}

The network we consider is a $K$ user interference network where transmitters $1,2,\cdots, K$ wish to send independent messages $W_1, W_2, \cdots, W_K$ to their respective destination decoders $1,2,\cdots, K$ over a heterogeneous two-hop network comprised of an intermediate stage of $K$ receive nodes. Each source is heard by three receive nodes in its neighborhood through noisy  wireless channels. Specifically, Source $k$ is heard by Receivers $k-1, k, k+1$. Destination decoder $k$ is able to access each of the received signals from Receivers $k$ and  $k+1$ through orthogonal, noiseless channels, which provides the destination decoder the opportunity for \emph{pairwise clustered decoding}, i.e., the joint processing of the two received signals to decode its desired message. Note that clustered processing is also of practical interest because, by definition, it limits the scope of cooperation. The heterogeneous nature of the network lies in the assumption of noise-free, orthogonal communication links from the first hop receivers to the final destination decoders, e.g., wired links of much higher capacity than the wireless channels of the first hop, so that the bottleneck remains the first hop. For ease of exposition -- i.e., in order to keep the network size small and to avoid edge effects at the same time -- we assume a ``wrap-around" model as the number of users is restricted to $K=4$ and the user indices are interpreted {\it modulo 4} as shown in Fig. \ref{fig:channelmodel} (User 4 can be alternatively thought of as User 0, but we prefer to number the users as 1,2,3,4). In the full paper \cite{fullpaper}, we show that the results reported here remain valid in an extended network of $K$ users without the wrap-around assumption, when $K$ is large enough to ignore edge-effects.

Fig. \ref{fig:channelmodel} shows the channel model with two kinds of links, indicated with solid red and dashed grey lines. We distinguish between two different connectivity settings.
\begin{enumerate}
\item {\bf Locally Connected: } The network of Fig. \ref{fig:channelmodel} is locally connected if and only if all the channel coefficients corresponding to the solid red lines take non-zero values while the channel coefficients for the dashed grey lines are set to zero.
\item {\bf Fully Connected: } The network of Fig. \ref{fig:channelmodel} is fully connected if and only if all the channel coefficients  take non-zero values.
\end{enumerate}

{\it Problem Statement: } Our goal in this paper is to explore the DoF of the network shown in Fig. \ref{fig:channelmodel} --- to introduce new achievability and outer bounding techniques that might find use in other settings as well, and to distill interesting insights into the impact of connectivity on the DoF of the cooperative network.

\subsection{Prior Work}
Of the vast amount of literature on cooperative wireless networks, the most closely related to this work  are references \cite{Gou_Jafar_MIMO, Levy_Shamai_ITA2009, Lapidoth_Shamai_ISIT2009, Shamai_ITA2010, Venu_ISIT2010}. In particular,  the  fully connected setting of Fig. \ref{fig:channelmodel} is shown to have DoF $= \frac{8}{3}$ in \cite{Venu_ISIT2010}. This  is identical to the 4-user SIMO (single input multiple output) interference channel setting without clustered decoding, where each user has one transmit and two receive antennas, and for which it is also established that the DoF = $8/3$ in \cite{Gou_Jafar_MIMO}.  While the outer bound and achievable precoding scheme extend from \cite{Gou_Jafar_MIMO} to \cite{Venu_ISIT2010} in a relatively straightforward manner, the proof of  achievability of $\frac{8}{3}$ DoF in the  fully connected clustered decoding setting of Fig. \ref{fig:channelmodel} is highly non-trivial, because of the strong spatial dependencies among desired and interference carrying channels (due to the sharing of receive antennas among decoders).  Moving beyond the fully connected setting, clustered decoding with \emph{local connectivity} is studied in \cite{Levy_Shamai_ITA2009, Lapidoth_Shamai_ISIT2009, Shamai_ITA2010}, under a variety of Wyner-type connectivity patterns and decoding-cluster formations. However,  the locally connected setting of Fig. \ref{fig:channelmodel} does not fall under any of the models addressed in these works. Furthermore, the tools used to obtain the DoF  inner bounds and outer bounds in \cite{Levy_Shamai_ITA2009, Lapidoth_Shamai_ISIT2009, Shamai_ITA2010} appear to be insufficient to find the DoF of the pairwise clustered decoding model of Fig. \ref{fig:channelmodel}.

\subsection{Contribution}
Our main contribution is an outer bound on the DoF of the locally connected setting of Fig. \ref{fig:channelmodel}. Specifically, we show that \emph{for every locally connected channel realization},  the interference network  shown in Fig. \ref{fig:channelmodel}, has { DoF  $ {\bf \leq}$  $\frac{\bf 12}{\bf 5}$}.

Further, we show that this is \emph{the best possible DoF outer bound} that could be valid for every locally connected channel realization, by constructing an explicit example where the outer bound is achieved.

\subsection{Significance}
We believe the contribution is interesting for two reasons.

First, as mentioned earlier in this section, the interplay between interference, cooperation and connectivity is of fundamental interest. This work illuminates some surprising aspects of the impact of connectivity on the DoF of cooperative interference networks. Specifically, our outer bound (DoF $\leq$  $\frac{12}{5}$)  for the locally connected setting (with only solid red channels present) is strictly \emph{lower} than the result of \cite{Venu_ISIT2010} that DoF $=$ $\frac{8}{3}$ for the fully connected setting. This is surprising because  the missing links in the locally connected setting apparently carry only undesired interference. Yet, removing these interference-carrying links \emph{reduces} the capacity, to the extent that even the DoF are strictly reduced. For instance, consider the dashed link from Transmitter 1 to Receiver 3. Since the signals from Receiver 3 are only available to Destination decoders 2 and 3, neither of which is interested in the message $W_1$ originating at Transmitter 1, it is somewhat surprising that removing this dashed channel reduces the DoF. Note that in the standard interference channel, removing interference carrying links cannot reduce capacity at all, much less reduce the DoF. A naive (and incorrect) explanation for the DoF loss could be that removing an interfering link reduces DoF because with two interfering links from the same undesired transmitter the decoder had the opportunity to cancel interference, which it cannot do with only one interfering link. Indeed, this is not the case. As we show in Section \ref{sec:insight}, there is no DoF loss from removing the interfering links in a similar setting where each decoder has access to two receive antennas with the same local connectivity pattern but without the spatial dependencies introduced by \emph{shared} receive antennas due to pairwise clustered decoding. Thus, the DoF outer bound  is  an illuminating indicator of the complex manner in which local connectivity impacts not only the capacity, but also the degrees of freedom of an interference network with collaborating nodes.

Second, the derivation of the outer bound itself follows a novel approach relative to prior work on related problems. While previously obtained DoF outer bounds  in \cite{Levy_Shamai_ITA2009, Lapidoth_Shamai_ISIT2009,Venu_ISIT2010}
%, Wigger_ITA09, Shamai_constrained_ITA10}
follow from a common genie-aided multiple-access channel argument, the outer bound derived here fundamentally relies on the sub-modularity property of entropy functions,  more commonly exploited in network coding converses \cite{Raymond_book, Blasiak_Kleinberg_Lubetzky}.
While the sub-modularity of  entropy functions is an elementary property by itself, it is the manner in which this property is applied that is quite insightful. Specifically, first, linear vector space dimension counting arguments are formulated based on the understanding of the role of interference alignment in this problem, and then these arguments are translated into information theoretic inequalities based on the equivalent sub-modularity properties of vector spaces and entropy functions\footnote{Due to space limitations, a description of the linear vector space alignment arguments that provide the intuition for the information theoretic converse is omitted here, and only the information theoretic converse is presented by itself. A detailed description of the insights behind this approach is relegated to the full paper \cite{fullpaper}.}. We expect that the general insights obtained in this work will be useful beyond the network of Fig. \ref{fig:channelmodel}.

\section{System model}
We begin by specifying the assumptions for the 4-user locally
connected interference network with pairwise clustered decoding, as
shown in Fig. \ref{fig:channelmodel} and described in the previous
section. We assume that channel coefficients corresponding to the
solid red links in Fig. \ref{fig:channelmodel} are allowed to vary
over time, can take \emph{any non-zero values}, and that global
channel state information (CSI) is available at all nodes. The
channel coefficients for all the dashed grey links in
Fig.\ref{fig:channelmodel} are zero, i.e., these channels are not
present.

The symbol received at Receiver $k$ over the $n^{th}$ channel use, $Y_k(n)$, is expressed as:
\begin{eqnarray*}
Y_k(n)&=&H_{kk}(n) X_k(n)+H_{k,k-1}(n)X_{k-1}(n)\\
&&+H_{k,k-1}X_{k-1}(n) + Z_k(n)
\end{eqnarray*}
where $H_{ij}(n)$ is the channel coefficient from Transmitter $j$ to first hop Receiver $i$,  $X_i$ is the symbol sent from Transmitter $i$ and $Z_k$ is the zero mean unit variance additive white Gaussian noise (AWGN) signal experienced by Receiver $k$. The input signals are assumed to have power $\rho$. Because we are interested primarily in DoF, we use $\rho$ and SNR interchangeably. The user index is interpreted in a cyclic wrap-around fashion.

There is pairwise clustered decoding so that the signals $Y_1^n,~Y_2^n$ are
jointly processed to decode message $W_1$; $Y_2^n,~Y_3^n$ to decode
$W_2$; $Y_3^n,~Y_4^n$ to decode $W_3$; and $Y_4^n,~Y_1^n$ to decode
$W_4$.

The capacity region $\mathcal {C}(\rho)$ of this network is a set of
achievable rate tuples ${\bf R}(\rho)
=\left(R_1(\rho),\ldots,R_4(\rho)\right)$ such that each user can simultaneously decode
its own message with arbitrarily small error
probability. The maximum sum rate of this
channel is defined as $R(\rho)=\max_{{\bf R}(\rho)\in \mathcal
{C}(\rho)}\sum_{k=1}^4 R_k(\rho)$. The capacity in the high SNR
regime can be characterized through DoF, i.e.,
DoF$=\lim_{\rho\rightarrow \infty}R(\rho)/\log\rho$.

Notation: We use the notation $o(x)$ to represent any function
$f(x)$ such that $\lim_{x\rightarrow \infty}\frac{f(x)}{x}=0$.

\section{DoF Outer Bound}

\begin{theorem}
The 4-user locally connected interference channel  has  DoF $\leq
\frac{12}{5}$ for all non-zero channel realizations.
\end{theorem}
{\it Proof}: Consider the achievable rate of User 1:
\begin{eqnarray}
nR_1&\!\!\!\!\leq \!\!\!\!&I(W_1;Y_1^n,Y_2^n)+o(n)\label{eqn:fano_ineq}\\
&\!\!\!\!\leq \!\!\!\!&I(W_1;Y_1^n,Y_2^n|W_2)+o(n)\label{eqn:cond_incr_rate}\\
&\!\!\!\!=\!\!\!\!&h(Y_1^n,Y_2^n|W_2)-h(Y_1^n,Y_2^n|W_1,W_2)+o(n)\\
&\!\!\!\!=\!\!\!\!&h(Y_1^n,Y_2^n|W_2)-n(R_3+R_4+o(\log\rho))+o(n)\ \ \ \ \label{eqn:invert_channel}\\
&\!\!\!\!\leq \!\!\!\!&h(Y_1^n|W_2)+h(Y_2^n|W_2)\notag\\
&\!\!\!\!&-n(R_3+R_4+o(\log\rho))+o(n)\ \ \ \label{eqn:chain_drop_cond}\\
&\!\!\!\!\leq \!\!\!\!&h(Y_1^n|W_1)+h(Y_1^n|W_4)-h(Y_1^n|W_1,W_4)\notag\\
&\!\!\!\!&+h(Y_2^n|W_2)-n(R_3+R_4+o(\log\rho))+o(n)\label{eqn:lemma1}\\
&\!\!\!\!=\!\!\!\!&h(Y_1^n|W_1)+h(Y_1^n|W_4)-n(R_2+o(\log\rho))\notag\\
&\!\!\!\!&+h(Y_2^n|W_2)-n(R_3+R_4+o(\log\rho))+o(n).\label{eqn:move_reconstr}
\end{eqnarray}
%\begin{small}
%\begin{eqnarray}
%nR_1&\!\!\!\!\leq \!\!\!\!&I(W_1;Y_1^n,Y_2^n)+o(n)\label{eqn:fano_ineq}\\
%&\!\!\!\!\leq \!\!\!\!&I(W_1;Y_1^n,Y_2^n|W_2)+o(n)\label{eqn:cond_incr_rate}\\
%&\!\!\!\!=\!\!\!\!&h(Y_1^n,Y_2^n|W_2)-h(Y_1^n,Y_2^n|W_1,W_2)+o(n)\\
%&\!\!\!\!=\!\!\!\!&h(Y_1^n,Y_2^n|W_2)-n(R_3+R_4+o(\log\rho))+o(n)\label{eqn:invert_channel}\\
%&\!\!\!\!\leq \!\!\!\!&h(Y_1^n|W_2)+h(Y_2^n|W_2)-n(R_3\!+\!R_4\!+\!o(\log\!\rho)\!)\!+\!o(n)\ \ \ \label{eqn:chain_drop_cond}\\
%&\!\!\!\!\leq \!\!\!\!&h(Y_1^n|W_1)+h(Y_1^n|W_4)-h(Y_1^n|W_1,W_4)\notag\\
%&\!\!\!\!&+h(Y_2^n|W_2)-n(R_3+R_4+o(\log\rho))+o(n)\label{eqn:lemma1}\\
%&\!\!\!\!=\!\!\!\!&h(Y_1^n|W_1)+h(Y_1^n|W_4)-n(R_2+o(\log\rho))\notag\\
%&\!\!\!\!&+h(Y_2^n|W_2)-n(R_3+R_4+o(\log\rho))+o(n).\label{eqn:move_reconstr}
%\end{eqnarray}
%\end{small}
Here, (\ref{eqn:fano_ineq}) follows from Fano's inequality.
(\ref{eqn:invert_channel}) follows from the invertibility of
upper/lower triangular channel matrices (regardless of the values of
the channel coefficients as long as they are all non-zero), which
implies that from the two output signals $Y_1^n, Y_2^n$, once we
remove the signals due to $W_1, W_2$, we obtain an interference-free
$2\times 2$ MIMO channel to transmitters $3, 4$ which can be
inverted to reconstruct the signals $X_3^n, X_4^n$ with noise
distortion that will depend on the channel coefficients but is
independent of SNR. From these noisy inputs, one can construct
signals statistically equivalent to the outputs $(Y_3^n, Y_4^n,
Y_1^n)$, again within noise tolerance that does not depend on SNR,
and from these outputs, possibly by reducing noise by an amount
independent of SNR, one can decode messages $W_3, W_4$. All these
operations only have an $o(\log({\rm SNR}))$ impact on rate, and so
we obtain $h(Y_1^n,
Y_2^n|W_1,W_2)=h(Y_3^n,Y_4^n,Y_1^n|W_1,W_2)+n~o(\log({\rm
SNR}))=n(R_3+R_4+o(\log({\rm SNR})))+o(n)$ as in
(\ref{eqn:invert_channel}). (\ref{eqn:chain_drop_cond}) follows from
chain rule of differential entropy and because dropping conditioning
cannot decrease differential entropy. (\ref{eqn:lemma1}) follows
from Lemma \ref{app:lemma1} shown in the Appendix. Note that it is
the use of Lemma \ref{app:lemma1} that invokes the sub-modularity
property of entropy functions. The intuition that Lemma
\ref{app:lemma1} should be applied in this manner comes from an
interference alignment perspective, to be elaborated upon in the
full paper.  Finally, (\ref{eqn:move_reconstr}) follows from the
observation that from $Y_1^n$, once all interference due to $W_1,
W_4$ is removed, we obtain an interference-free AWGN channel to
transmitter $2$, from which $W_2$ can be decoded subject to noise
reduction operations that only have an $o(\log({\rm SNR}))$ impact.
Therefore $h(Y_1^n|W_1,W_4)=n(R_2+o(\log({\rm SNR})))+o(n)$ as in
(\ref{eqn:move_reconstr}).

What we have so far is the first set of bounds:
\begin{eqnarray}
n(R_1+R_2+R_3+R_4+o(\log\rho))\leq\ \ \ \ \ \ \ \ \ \ \ \ \ \ \ \ \notag\\
\ \ \ \ \ \ \ \ \ h(Y_1^n|W_1)+h(Y_1^n|W_4)+h(Y_2^n|W_2)+o(n)\label{eqn:rearrange_reconstr}\\
n(R_2+R_3+R_4+R_1+o(\log\rho))\leq\ \ \ \ \ \ \ \ \ \ \ \ \ \ \ \ \notag\\
\ \ \ \ \ \ \ \ \ h(Y_2^n|W_2)+h(Y_2^n|W_1)+h(Y_3^n|W_3)+o(n)\label{eqn:first_set_2}\\
n(R_3+R_4+R_1+R_2+o(\log\rho))\leq\ \ \ \ \ \ \ \ \ \ \ \ \ \ \ \ \notag\\
\ \ \ \ \ \ \ \ \ h(Y_3^n|W_3)+h(Y_3^n|W_2)+h(Y_4^n|W_4)+o(n)\\
n(R_4+R_1+R_2+R_3+o(\log\rho))\leq\ \ \ \ \ \ \ \ \ \ \ \ \ \ \ \ \notag\\
\ \ \ \ \ \ \ \ \
h(Y_4^n|W_4)+h(Y_4^n|W_3)+h(Y_1^n|W_1)+o(n)\label{eqn:first_set_K}
\end{eqnarray}
%\begin{small}
%\begin{eqnarray}
%n(R_1+R_2+R_3+R_4+o(\log\rho))\leq\ \ \ \ \ \ \ \ \ \ \ \ \ \ \ \ \ \ \notag\\
%\ \ \ \ \ \ \ \ \ \ \ \ \ \ \ h(Y_1^n|W_1)+h(Y_1^n|W_4)+h(Y_2^n|W_2)+o(n)\label{eqn:rearrange_reconstr}\\
%n(R_2+R_3+R_4+R_1+o(\log\rho))\leq\ \ \ \ \ \ \ \ \ \ \ \ \ \ \ \ \ \ \notag\\
%\ \ \ \ \ \ \ \ \ \ \ \ \ \ \ h(Y_2^n|W_2)+h(Y_2^n|W_1)+h(Y_3^n|W_3)+o(n)\label{eqn:first_set_2}\\
%n(R_3+R_4+R_1+R_2+o(\log\rho))\leq\ \ \ \ \ \ \ \ \ \ \ \ \ \ \ \ \ \ \notag\\
%\ \ \ \ \ \ \ \ \ \ \ \ \ \ \ h(Y_3^n|W_3)+h(Y_3^n|W_2)+h(Y_4^n|W_4)+o(n)\\
%n(R_4+R_1+R_2+R_3+o(\log\rho))\leq\ \ \ \ \ \ \ \ \ \ \ \ \ \ \ \ \ \ \notag\\
%\ \ \ \ \ \ \ \ \ \ \ \ \ \ \
%h(Y_4^n|W_4)+h(Y_4^n|W_3)+h(Y_1^n|W_1)+o(n)\label{eqn:first_set_K}
%\end{eqnarray}
%\end{small}
where (\ref{eqn:rearrange_reconstr}) is obtained by rearranging the
terms in (\ref{eqn:move_reconstr}) and the remaining inequalities
(\ref{eqn:first_set_2}) to (\ref{eqn:first_set_K}) follow by
symmetry by simply cyclically advancing the user indices.

Next we  obtain the second set of bounds:
\begin{eqnarray}
nR_1&\!\!\!\!\leq\!\!\!\!&I(Y_1^n,Y_2^n;W_1)+o(n)\\
&\!\!\!\!=\!\!\!\!&h(Y_1^n,Y_2^n)-h(Y_1^n,Y_2^n|W_1)+o(n)\\
&\!\!\!\!=\!\!\!\!&h(Y_1^n,Y_2^n)-I(Y_1^n,Y_2^n;W_2|W_1)\notag\\
&\!\!\!\!&-h(Y_1^n,Y_2^n|W_1,W_2)+o(n)\label{eqn:alternative_R1}\\
&\!\!\!\!\leq\!\!\!\!&h(Y_1^n,Y_2^n)-I(Y_1^n;W_2|W_1)\notag\\
&\!\!\!\!&-n(R_3+R_4+o(\log\rho))+o(n)\\
&\!\!\!\!=\!\!\!\!&h(Y_1^n,Y_2^n)-h(Y_1^n|W_1)+h(Y_1^n|W_1,W_2)\notag\\
&\!\!\!\!&-n(R_3+R_4+o(\log\rho))+o(n)\\
&\!\!\!\!=\!\!\!\!&h(Y_1^n,Y_2^n)-h(Y_1^n|W_1)+n(R_4+o(\log\rho))\ \ \ \ \notag\\
&\!\!\!\!&-n(R_3+R_4+o(\log\rho))+o(n)\label{eq:explain}\\
&\!\!\!\!\leq\!\!\!\!&2n(\log\rho+o(\log\rho))-h(Y_1^n|W_1)\notag\\
&\!\!\!\!&-n(R_3+o(\log\rho))+o(n)\label{eqn:for_set2_11}
\end{eqnarray}
%\begin{small}
%\begin{eqnarray}
%nR_1\!\!&\!\!\!\!\leq\!\!\!\!&\!\!I(Y_1^n,Y_2^n;W_1)+o(n)\\
%&\!\!\!\!=\!\!\!\!&\!\!h(Y_1^n,Y_2^n)-h(Y_1^n,Y_2^n|W_1)+o(n)\\
%&\!\!\!\!=\!\!\!\!&\!\!h(Y_1^n\!,\!Y_2^n\!)\!-\!I(Y_1^n\!,\!Y_2^n\!;\!W_2|W_1\!)\!-\!h(Y_1^n\!,\!Y_2^n|W_1,\!W_2\!)\!+\!o(\!n\!)\ \ \ \ \\
%&\!\!\!\!\leq\!\!\!\!&\!\!h(Y_1^n\!,\!Y_2^n\!)\!-\!I(Y_1^n\!;\!W_2|W_1\!)\!-\!n(R_3\!+\!R_K\!+\!o(\log\!\rho)\!)\!+\!o(\!n\!)\ \ \ \\
%&\!\!\!\!=\!\!\!\!&\!\!h(Y_1^n,Y_2^n)-h(Y_1^n|W_1)+h(Y_1^n|W_1,W_2)\notag\\
%&\!\!\!\!&-n(R_3+R_K+o(\log\rho))+o(n)\\
%&\!\!\!\!=\!\!\!\!&\!\!h(Y_1^n,Y_2^n)-h(Y_1^n|W_1)+n(R_K+o(\log\rho))\notag\\
%&\!\!\!\!&-n(R_3+R_K+o(\log\rho))+o(n)\\
%&\!\!\!\!\leq\!\!\!\!&\!\!2n\log\!\rho\!+\!o(\log\!\rho)\!-\!h(Y_1^n|W_1)\!-\!n(R_3\!+\!o(\log\rho))\!+\!o(n)\label{eqn:for_set2_11}
%\end{eqnarray}
%\end{small}
and again starting from (\ref{eqn:alternative_R1}) in an
alternative fashion:
\begin{eqnarray}
nR_1&\!\!\!\!\leq\!\!\!\!&h(Y_1^n,Y_2^n)-I(Y_1^n,Y_2^n;W_2|W_1)\notag\\
&\!\!\!\!&-h(Y_1^n,Y_2^n|W_1,W_2)+o(n)\\
&\!\!\!\!\leq\!\!\!\!&h(Y_1^n,Y_2^n)-I(Y_2^n;W_2|W_1)\notag\\
&\!\!\!\!&-n(R_3+R_4+o(\log\rho))+o(n)\\
&\!\!\!\!=\!\!\!\!&h(Y_1^n,Y_2^n)-h(Y_1^n|W_1)+h(Y_2^n|W_1,W_2)\notag\\
&\!\!\!\!&-n(R_3+R_4+o(\log\rho))+o(n)\\
&\!\!\!\!=\!\!\!\!&h(Y_1^n,Y_2^n)-h(Y_1^n|W_1)+n(R_3+o(\log\rho))\ \ \ \ \notag\\
&\!\!\!\!&-n(R_3+R_4+o(\log\rho))+o(n)\label{eq:explain2}\\
&\!\!\!\!\leq\!\!\!\!&2n(\log\rho+o(\log\rho))-h(Y_2^n|W_1)\notag\\
&\!\!\!\!&-n(R_4+o(\log\rho))+o(n).\label{eqn:for_set2_12}
\end{eqnarray}

{\it Remark: } In arriving at (\ref{eq:explain}) we use the
substitution $h(Y_1^n|W_1,W_2)=n(R_4+o(\log({\rm SNR})))+o(n)$. This
is derived explicitly in the Appendix as Lemma \ref{lemma:sub}.
Similar substitution is made in arriving at (\ref{eq:explain2}) as
well. These substitutions make use of the assumption that the
mapping from messages $W_i$ to the codewords $X_i^n$ is
deterministic and invertible. It is evident that this assumption
does not incur any loss of generality in our setting, where it can
be argued that the best deterministic codebook will perform at least
as well as a randomized coding scheme.

%\begin{small}
%\begin{eqnarray}
%nR_1&\!\!\!\!\leq\!\!\!\!&h(Y_1^n,Y_2^n)-I(Y_1^n,Y_2^n;W_2|W_1)-h(Y_1^n,Y_2^n|W_1,W_2)+o(n)\\
%&\!\!\!\!\leq\!\!\!\!&h(Y_1^n,Y_2^n)-I(Y_2^n;W_2|W_1)-n(R_3+R_K+o(\log\rho))+o(n)\\
%&\!\!\!\!=\!\!\!\!&h(Y_1^n,Y_2^n)-h(Y_1^n|W_1)+h(Y_2^n|W_1,W_2)\notag\\
%&\!\!\!\!&-n(R_3+R_K+o(\log\rho))+o(n)\\
%&\!\!\!\!=\!\!\!\!&h(Y_1^n,Y_2^n)-h(Y_1^n|W_1)+n(R_3+o(\log\rho))\notag\\
%&\!\!\!\!&-n(R_3+R_K+o(\log\rho))+o(n)\\
%&\!\!\!\!\leq\!\!\!\!&2n\log\rho+o(\log\rho)-h(Y_2^n|W_1)-n(R_K+o(\log\rho))+o(n).\label{eqn:for_set2_12}
%\end{eqnarray}
%\end{small}
Thus, we have the second set of inequalities:
\begin{eqnarray}
h(Y_1^n|W_1)&\!\!\!\!\leq \!\!\!\!&n(2\log\rho-R_1-R_3+o(\log\rho))+o(n)\label{eqn:second_set_11}\ \ \ \ \ \ \\
h(Y_2^n|W_1)&\!\!\!\!\leq \!\!\!\!&n(2\log\rho-R_1-R_4+o(\log\rho))+o(n)\label{eqn:second_set_12}\\
h(Y_2^n|W_2)&\!\!\!\!\leq \!\!\!\!&n(2\log\rho-R_2-R_4+o(\log\rho))+o(n)\label{eqn:second_set_21}\ \ \ \ \ \ \\
h(Y_3^n|W_2)&\!\!\!\!\leq \!\!\!\!&n(2\log\rho-R_2-R_1+o(\log\rho))+o(n)\\
h(Y_3^n|W_3)&\!\!\!\!\leq \!\!\!\!&n(2\log\rho-R_3-R_1+o(\log\rho))+o(n)\\
h(Y_4^n|W_3)&\!\!\!\!\leq \!\!\!\!&n(2\log\rho-R_3-R_2+o(\log\rho))+o(n)\\
h(Y_4^n|W_4)&\!\!\!\!\leq \!\!\!\!&n(2\log\rho-R_4-R_2+o(\log\rho))+o(n)\\
h(Y_1^n|W_4)&\!\!\!\!\leq
\!\!\!\!&n(2\log\rho-R_4-R_3+o(\log\rho))+o(n)\label{eqn:second_set_K2}
\end{eqnarray}
%\begin{small}
%\begin{eqnarray}
%h(Y_1^n|W_1)&\!\!\!\!\leq \!\!\!\!&2n\log\rho-n(R_1+R_3)+o(\log\rho)+o(n)\label{eqn:second_set_11}\ \ \ \ \\
%h(Y_2^n|W_1)&\!\!\!\!\leq \!\!\!\!&2n\log\rho-n(R_1+R_4)+o(\log\rho)+o(n)\label{eqn:second_set_12}\\
%h(Y_2^n|W_2)&\!\!\!\!\leq \!\!\!\!&2n\log\rho-n(R_2+R_4)+o(\log\rho)+o(n)\label{eqn:second_set_21}\\
%h(Y_3^n|W_2)&\!\!\!\!\leq \!\!\!\!&2n\log\rho-n(R_2+R_1)+o(\log\rho)+o(n)\\
%h(Y_3^n|W_3)&\!\!\!\!\leq \!\!\!\!&2n\log\rho-n(R_3+R_1)+o(\log\rho)+o(n)\\
%h(Y_4^n|W_3)&\!\!\!\!\leq \!\!\!\!&2n\log\rho-n(R_3+R_2)+o(\log\rho)+o(n)\\
%h(Y_4^n|W_4)&\!\!\!\!\leq \!\!\!\!&2n\log\rho-n(R_4+R_2)+o(\log\rho)+o(n)\\
%h(Y_1^n|W_4)&\!\!\!\!\leq
%\!\!\!\!&2n\log\rho-n(R_4+R_3)+o(\log\rho)+o(n)\label{eqn:second_set_K2}
%\end{eqnarray}
%\end{small}
where (\ref{eqn:second_set_11}), (\ref{eqn:second_set_12}) are
rearranged forms of (\ref{eqn:for_set2_11}),
(\ref{eqn:for_set2_12}), respectively, and the remaining
inequalities (\ref{eqn:second_set_21}) to (\ref{eqn:second_set_K2})
are the symmetric versions of (\ref{eqn:second_set_11}),
(\ref{eqn:second_set_12}) obtained by cyclically advancing the
indices.

Substituting the right-hand side of the inequalities
(\ref{eqn:second_set_11}) to (\ref{eqn:second_set_K2}) wherever the
corresponding left-hand side appears in the inequalities
(\ref{eqn:rearrange_reconstr}) to (\ref{eqn:first_set_K}), and
adding up all the inequalities, we obtain:
\begin{eqnarray}
4n(R\!+\!o(\log\!\rho))\!\leq\!12(2n\log\!\rho)\!-\!6n(R\!+\!o(\log\!\rho))\!+\!o(n)
\end{eqnarray}
where $R=R_1+R_2+R_3+R_4$. Rearranging terms, dividing by $n$ and
taking the limit $n\rightarrow \infty$ we have:
\begin{eqnarray}
4R\leq24\log\rho-6R+o(\log\rho).
\end{eqnarray}
Now applying the limit $\rho\rightarrow \infty$ on the sum rate
outer bound, we have the DoF outer bound:
\begin{eqnarray}
{\rm
DoF}=\lim_{\rho\rightarrow\infty}\frac{R}{\log\rho}\leq\frac{12}{5}.
\end{eqnarray}
\hfill\QED

\section{A Locally Connected Example with $12/5$ DoF}

The DoF outer bound shown in the previous section holds for all
locally connected channel realizations. In this section we prove that a better outer bound is not possible in the same sense, by providing
an example of a locally connected channel realization where  $12/5$ DoF are achieved.

For the locally connected 4-user interference channel we consider in
this article, consider a 5 symbol extension. So each transmitter and
receiver has access to a $5\times 5$ MIMO channel over 5 time lots.
Note that these MIMO channels have a diagonal structure due to the
nature of symbol extensions. In order to achieve $12/5$ DoF, each
user needs to send 3 symbols over 5 times slots. We artificially
construct the channel matrices between each transmitter $T_j$ and
each receiver $R_i$ as follows. The channel matrix from $T_j$ to
$R_i$ is denoted as ${\bf H}_{ij}$. Here ${\bf I}$, ${\bf O}$ stand
for the identity and zero matrices, respectively, and the diagonal
matrix ${\bf G}={\rm diag}([1~2~3~4~5])$.

\begin{figure}[!h] \centering \vspace{-0.05in}
\includegraphics[width=2in]{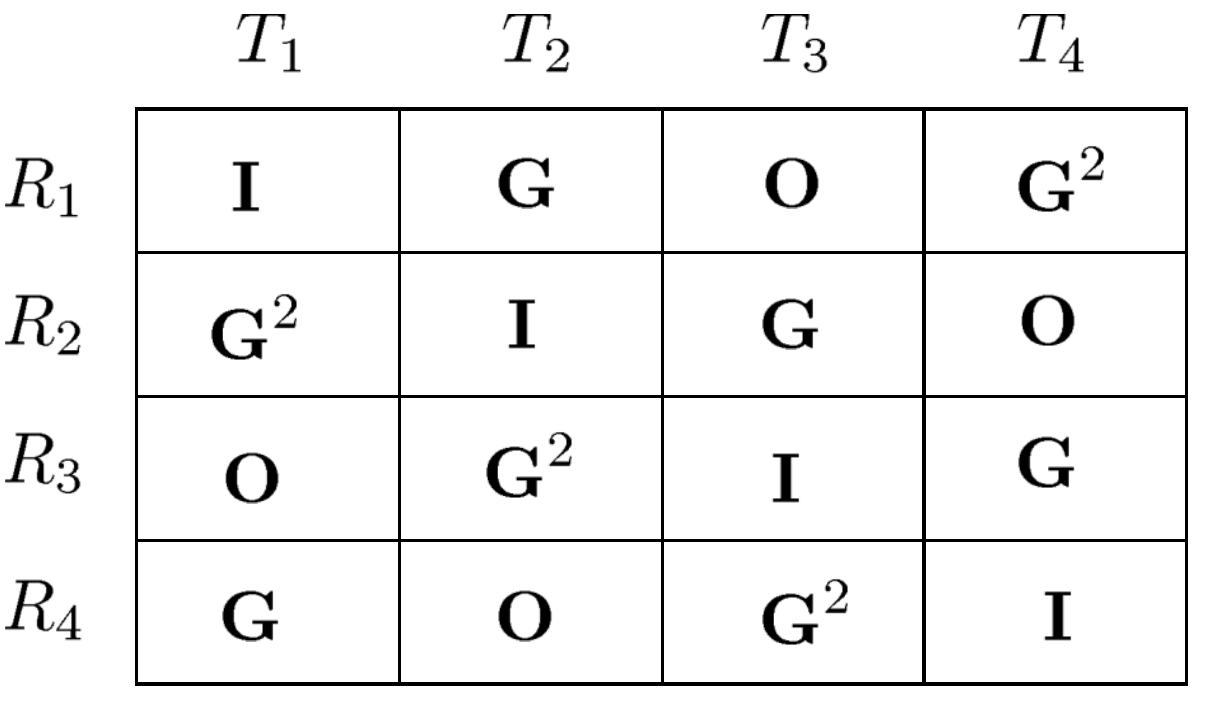}
\vspace{-0.15in} \caption{The channel matrices for the artificial
example} \label{fig:toy_model} \vspace{-0.05in}
\end{figure}

Each transmitter uses the
same three beamforming vectors ${\bf w}$, ${\bf G}{\bf w}$ and ${\bf
G}^2{\bf w}$ to send its three symbols $u^{[k]}_1$, $u^{[k]}_2$ and
$u^{[k]}_3$ where ${\bf w}=\left[1~1~1~1~1\right]^T$ and $k$ is the
user index. In order to see how this scheme works, let us consider,
without loss of generality, the received signal vector at user 1.

Besides its own output $\overline{Y}_1$ over 5 time slots, user 1
can also access its neighboring output $\overline{Y}_2$.
Thus, user 1 is able to see a 10 dimensional space. The received
signal of user 1 over the 5 time slots, ${\bf
y}^{[1]}=\left[\overline{Y}_1^T~\overline{Y}_2^T\right]^T$, is given
by:
\begin{small}
\begin{eqnarray}
{\bf y}^{[1]}\!=\!\underbrace{\!\left[\!\!\!\begin{array}{c}{\bf
H}_{11}\\{\bf H}_{21}\end{array}\!\!\!\right]\!{\bf x}^{[1]}\!}_{\rm
desired~ signal}\!+\!\underbrace{\!\left[\!\!\!\begin{array}{c}{\bf
H}_{12}\\{\bf H}_{22}\end{array}\!\!\!\right]\!{\bf
x}^{[2]}\!+\!\left[\!\!\!\begin{array}{c}{\bf H}_{13}\\{\bf
H}_{23}\end{array}\!\!\!\right]\!{\bf
x}^{[3]}\!+\!\left[\!\!\!\begin{array}{c}{\bf H}_{14}\\{\bf
H}_{24}\end{array}\!\!\!\right]\!{\bf x}^{[4]}\!}_{\rm
interference}+{\bf z}^{[1]}
\end{eqnarray}
\end{small}
where ${\bf z}^{[1]}$ is the noise vector and ${\bf x}^{[k]}$ is the
$5\times 1$ transmit signal vector of user $k$ which is given by:
\begin{eqnarray}
{\bf x}^{[k]}=\underbrace{\left[{\bf w}~{\bf G}{\bf w}~{\bf G}^2{\bf
w}\right]}_{\triangleq {\bf
B}^{[k]}}\underbrace{\left[u^{[k]}_1~u^{[k]}_2~u^{[k]}_3\right]^T}_{\triangleq
{\bf u}^{[k]}}.\label{eqn:toy_tr_signal}
\end{eqnarray}

In order to preserve a 3 dimensional space for user 1's desired
signals, we need to align the 9 interference streams (3 per
interferer) into a 7 dimensional space. In other words, we need to
ensure that the rank of the following matrix ${\bf H}_{\bf I}$,
whose column vectors span the space occupied by the interference, is
no larger than 7:
\begin{eqnarray}
{\bf H}_{\bf I}=\left[\begin{array}{ccc}{\bf H}_{12}{\bf
B}^{[2]}&{\bf H}_{13}{\bf B}^{[3]}&{\bf H}_{14}{\bf B}^{[4]}\\{\bf
H}_{22}{\bf B}^{[2]}&{\bf H}_{23}{\bf B}^{[3]}&{\bf H}_{24}{\bf
B}^{[4]}\end{array}\right].
\end{eqnarray}
Since the numerical values of all quantities are known, this is easily verified.
It turns out that the second and third column vectors align in the
space spanned by the fourth, fifth, seventh and eighth column
vectors.

What remains to be shown is that the 3 dimensions carrying the three
desired symbols are linearly independent with the remaining 7 column
vectors carrying interference. For this, we prove the
following matrix consisting of the ten column vectors (eliminating
the second and third column vectors of ${\bf H}_{\bf I}$ which align with
the remaining interference) has full rank.
\begin{small}
\begin{eqnarray}
{\bf H}=\left[\!\!\begin{array}{cccccccc}{\bf H}_{11}{\bf
B}^{[1]}\!&\!{\bf G}{\bf w}\!\!&\!\!{\bf 0}\!\!&\!\!{\bf
0}\!\!&\!\!{\bf 0}\!\!&\!\!{\bf G}^2{\bf w}\!\!&\!\!{\bf G}^3{\bf
w}\!\!&\!\!{\bf G}^4{\bf w}\\{\bf H}_{21}{\bf B}^{[1]}\!&\!{\bf
w}\!\!&\!\!{\bf G}{\bf w}\!\!&\!\!{\bf G}^2{\bf w}\!\!&\!\!{\bf
G}^3{\bf w}\!\!&\!\!{\bf 0}\!\!&\!\!{\bf 0}\!\!&\!\!{\bf
0}\end{array}\!\!\right].
\end{eqnarray}
\end{small}
This is also easily verified by explicitly evaluating the determinant of this channel matrix.
Similar analysis can be carried out to user 2, 3 and 4 due to
cyclical symmetry in the construction of channels.

\section{Channel Connectivity and Cooperation}\label{sec:insight}

As mentioned briefly in the introduction, we have shown
that for the network shown in Fig. \ref{fig:channelmodel}, the locally connected setting has strictly smaller DoF than the fully connected setting, even though the additional channel coefficients in the fully connected setting apparently carry only undesired interference.

\begin{figure}[!h] \centering
\includegraphics[width=3.1in]{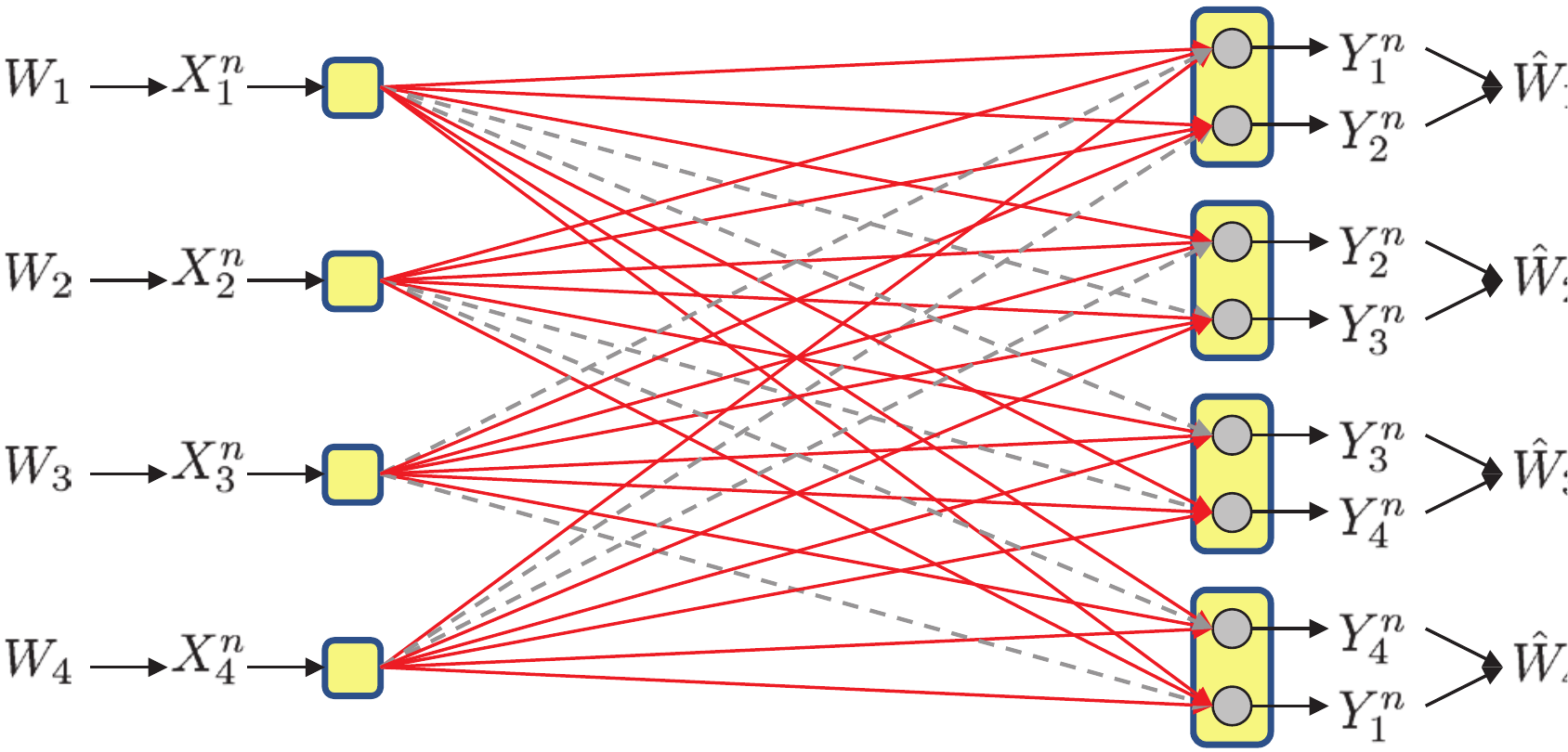}
\caption{4 User Interference Channel with Cooperative
Receivers}\vspace{-0.1in} \label{fig:connectivity}
\end{figure}

In order to take a more refined look at this phenomenon, we investigate if the DoF loss is
caused by the pairwise clustered cooperation at the receiver side, or
local channel links connectivity. In Fig.\ref{fig:connectivity}, we
show an equivalent representation of the network in
Fig.\ref{fig:channelmodel} using the style of SIMO interference
channel, but introducing spatial dependencies between some channel
coefficients. Specifically, the channel coefficient associated with
the second antenna of Receiver $k$ is identical to that associated with
the first antenna of Receiver $k+1$.

%\begin{figure}[!h] \centering
%\includegraphics[width=2.8in]{connectivity_single}
%\caption{The channel links associated with user 1}\vspace{-0.1in}
%\label{fig:connectivity_single}
%\end{figure}

First, let us consider the impact of clustered decoding alone, by removing the locally connected
assumption. Consider the network in
Fig.\ref{fig:connectivity} in the fully connected setting, i.e., coefficients of
all solid and dashed links are generic and non-zero. In this case, whether each decoder has access
to two independent receive antennas (not shared with any other decoder), or the decoders share
receive antennas as in Fig. \ref{fig:channelmodel}, in both cases DoF = 8/3. Thus,  the spatial dependencies between channel
coefficients induced by clustered decoding (sharing of received antennas between decoders) do not not affect the DoF of this channel.

Second, we consider the role of local connectivity but without the spatial dependencies
introduced by clustered decoding. Let us
remove the dashed links in Fig.\ref{fig:connectivity}. Again, if there are no spatial
dependencies between any channel links,  then it forms a classical
SIMO interference channel with dashed links disconnected. Interestingly, we can show that for this network, the DoF is still equal to $8/3$, in spite of the local connectivity.

%Intuitively, this is because we can align the signal vectors of user 2 into the space spanned by the signal vectors from user 3 and user 4. The independence of direct links from transmitter 1 and interference-carrying links from transmitters 2, 3 and 4 is significant here. Thus, the achievable schemes created in \cite{Gou_Jafar_MIMO} can be applied directly here.

Thus, we find that the spatial dependencies caused by clustered decoding do not reduce the
DoF if the channel is fully connected. Similarly, the local connectivity does not reduce the DoF if
there are no spatial dependencies caused by clustered decoding (shared receive antennas across decoders). That is, individually, neither
clustered decoding, nor local connectivity causes a loss of DoF. However, as shown by the outer bound,
when taken together, the spatial dependencies caused by clustered decoding in conjunction with the
local connectivity, translate into a DoF loss.

\section{Conclusion}
We derived a new information theoretic outer bound on the degrees of
freedom (DoF) for a 4-user locally connected interference channel
with pairwise clustered decoding. Interestingly, we found that
removing interference-carrying links decreases the DoF. The outer
bound derivation incorporates novel elements and insights that may
be useful beyond the problem considered in this work.

The DoF with generic channels (i.e., in the almost surely sense)
remain open for the locally connected 4-user interference channel
with pairwise clustered decoding. For more than $4$ users and
generic channels, the DoF remain open for both the fully connected
and locally connected cases with pairwise clustered decoding,
although it is known that there is a loss of DoF relative to the
corresponding SIMO interference channel for the fully connected
case. Interestingly, for more than 4 users there is a gap between
the best outer bound valid for all non-zero channel realizations in
the fully connected case (which continues to be 2/3 DoF per user and
can be shown to be achievable for certain realizations) and smaller
outer bounds that can be shown to be valid for \emph{almost} all
channel coefficients. For example,  the fully connected $K$ user
interference network  with $M$ sized clustered decoding has a
straightforward DoF outer bound of $1/2+(M-1)/K$ per user (based on
the usual multiple-access type outer bounding arguments)  that is
valid for almost all values of channel coefficients. Interestingly,
this shows that the DoF per user in the fully connected setting lose
all benefits of clustered decoding as the network size becomes
large, for almost all channel realizations.  Similar DoF
characterizations for generic channel realizations in the locally
connected setting remain a challenging open problem.

{\it Acknowledgment: }
%\section*{Acknowledgment}
The work of Chenwei Wang and Syed Jafar is supported by NSF. The
work of Shlomo Shamai is supported by the Israel Science Foundation.
% that's all folks

%The DoF analysis sheds interesting insights into the role of clustered decoding and local connectivity in the DoF loss relative to the fully connected setting with independent received antennas. such as this, and the novel alignment schemes as well as new outer bounds needed to account for  shared antennas, make this problem of great interest. The reciprocal setting of cooperative transmitters is similarly interesting.

\newpage

\appendix
\section{Appendix}
\begin{lemma}\label{app:lemma1}
\begin{eqnarray}
h(Y_1^n|W_2)\!+\!h(Y_1^n|W_1,W_4)\!\leq
\!h(Y_1^n|W_1)\!+\!h(Y_1^n|W_4)
\end{eqnarray}
\end{lemma}
{\it Proof}:
\begin{eqnarray}
&\!\!&h(Y_1^n|W_2)+h(Y_1^n|W_1,W_4)\\
&\!\!=\!\!&2h(Y_1^n)-I(W_2;Y_1^n)-I(Y_1;W_1,W_4)\\
&\!\!\leq\!\!&2h(Y_1^n)-I(Y_1;W_1,W_4)\\
&\!\!=\!\!&2h(Y_1^n)-I(Y_1^n;W_1)-I(Y_1;W_4|W_1)\ \ \ \ \label{eqn:app_cond_incr_rate}\\
&\!\!\leq\!\!&2h(Y_1^n)-I(Y_1^n;W_1)-I(Y_1;W_4)\label{eqn:app_incr_rate}\\
&\!\!=\!\!&h(Y_1^n|W_1)+h(Y_1^n|W_4)
\end{eqnarray}
where (\ref{eqn:app_incr_rate}) follows from
(\ref{eqn:app_cond_incr_rate}) because $I(A;B|C)\geq I(A;B)$ when
$A$ is independent of $C$. \hfill\QED
\\
\begin{lemma}\label{lemma:sub}
\begin{eqnarray}
h(Y_1^n|W_1,W_2)=n(R_4+o(\log({\rm SNR})))+o(n)
\end{eqnarray}
\end{lemma}
{\it Proof}:
We use $H_{ij}$ to denote  the channel coefficient
between Receiver $i$ and Transmitter $j$. $Z_i$ is the additive
white Gaussian noise term at Receiver $i$.
\begin{eqnarray}
nR_4&\!\!\!\!=\!\!\!\!& H(W_4)\\
&\!\!\!\!=\!\!\!\!&I(W_4; H_{14} X_4^n + Z_1^n, H_{44}X_4^n+Z_4^n)\notag\\
&\!\!\!\!&+H(W_4|H_{14}X_4^n+Z_1^n,H_{44}X_4^n+Z_4^n)\\
&\!\!\!\!=\!\!\!\!&I(W_4; H_{14} X_4^n + Z_1^n, H_{44}X_4^n+Z_4^n)+o(n)\label{eq:Fanosub}\\
&\!\!\!\!=\!\!\!\!&I(W_4; H_{14} X_4^n + Z_1^n)\notag\\
&\!\!\!\!&+I(W_4;  H_{44}X_4^n+Z_4^n|H_{14} X_4^n + Z_1^n)+o(n)\\
&\!\!\!\!=\!\!\!\!&I(W_4; H_{14} X_4^n + Z_1^n)+n~o(\log({\rm SNR}))+o(n)\\
&\!\!\!\!=\!\!\!\!&I(W_4; Y_1^n | W_1, W_2) + n ~o(\log({\rm SNR})) + o(n)\label{eq:detcod}\\
&\!\!\!\!=\!\!\!\!& h(Y_1^n|W_1,W_2)-h(Y_1^n|W_1,W_2,W_4)\notag\\
&\!\!\!\!&+ n ~o(\log({\rm SNR})) + o(n)\\
&\!\!\!\!=\!\!\!\!&h(Y_1^n|W_1,W_2)-h(Z_1^n)+ n ~o(\log({\rm SNR})\!)\!+\!o(n\!)\ \ \ \ \ \\
&\!\!\!\!=\!\!\!\!&h(Y_1^n|W_1,W_2)+ n ~o(\log({\rm SNR})) + o(n)
\end{eqnarray}
where (\ref{eq:Fanosub}) follows from Fano's inequality,
(\ref{eq:detcod}) follows from the assumption of deterministic and
invertible mapping from messages to codewords. Note that
$o(\log({\rm SNR}))$ terms in this set of equations can be described
more tightly as $O(1)$ terms, i.e., bounded terms that do not
increase with ${\rm SNR}$. However, for our purpose, since we are
primarily interested in the DoF, it suffices to highlight only their
$o(\log({\rm SNR}))$ character. \hfill\QED

\end{document}